\documentclass[reprint,showpacs,amsmath,amssymb,aps,superscriptaddress,floatfix,longbibliography,nofootinbib]{revtex4-2}
\usepackage{amsthm}
\usepackage{amsmath, mathtools}
\usepackage[utf8]{inputenc}
\usepackage[english]{babel}
\usepackage{amsmath}
\usepackage{graphicx}		
\usepackage{natbib}
\usepackage{textcomp}
\usepackage{gensymb}
\usepackage[usenames,dvipsnames,svgnames,table]{xcolor}
\usepackage[hidelinks,colorlinks=false,urlcolor=Cerulean,citecolor=black]{hyperref}
\usepackage{siunitx}
\usepackage{mathrsfs}
\usepackage{multirow}
\usepackage{bm}
\usepackage{xfrac}
\usepackage{comment}
\usepackage[normalem]{ulem}

\renewcommand{\i}{\mathrm{i}}

\DeclareMathOperator{\N}{\mathbb{N}}

\begin{document}

\title{Delay coordinates synchronization and induces abrupt transition in excitable networks}

\author{Bruno R. R. Boaretto}
\thanks{Co-first authors}
\affiliation{Institute of Science and Technology, Universidade Federal de São Paulo (UNIFESP), São José dos Campos SP, Brazil}
\altaffiliation[Contact email: ]{bruno.boaretto@unifesp.br}

\author{Kalel L. Rossi}
\thanks{Co-first authors}
\affiliation{Cellular Computations and Learning, Max Planck Institute for Neurobiology of Behavior, Bonn, Germany}

\author{Lyle E. Muller}
\affiliation{Department of Mathematics, Western University, London ON, Canada}
\affiliation{Fields Lab for Network Computation, Fields Institute, Toronto ON, Canada}

\author{Elbert E. Macau}
\thanks{Co-senior authors}
\affiliation{Institute of Science and Technology, Universidade Federal de São Paulo (UNIFESP), São José dos Campos SP, Brazil}

\author{Roberto C. Budzinski}
\thanks{Co-senior authors}
\affiliation{Department of Neuroscience, University of Lethbridge, Lethbridge AB, Canada}
\affiliation{Fields Lab for Network Computation, Fields Institute, Toronto ON, Canada}

\begin{abstract}
Neuronal communication is inherently time-delayed, due to the finite speed of signal propagation. Although often considered challenging or disruptive, such time delays can also endow neural circuits with useful capabilities. Here, we show that delays in excitatory connections between excitable neurons coordinate their synchronization patterns by creating self-sustained oscillations that may be out-of-phase or in-phase. The emergence of these oscillations leads to an abrupt, explosive, transition to in-phase synchronized regimes due to small changes in connection strength or time-delay. We describe the mechanism underlying these phenomena as an interaction between the neuron's excitable dynamics and the delay in signal transmission, explaining many aspects of how the oscillations emerge. We show this phenomenon in different network connectivities, neuronal models, with and without excitation, with and without noise, highlighting the generality of the mechanism.
\end{abstract}

\maketitle

Time delays are ubiquitous in interacting systems, arising from finite signal propagation speeds, synaptic transmission times, and processing latencies. In the brain, delays are intrinsic to the connectivity at multiple scales, from local cortical circuits to long-range cortico-cortical interactions \cite{girard2001feedforward,swadlow2012axonal,muller2018cortical}. These temporal lags influence neural activity \cite{muller2018cortical,takahashi2015large}, shape resting-state dynamics \cite{deco2009key}, may contribute to synaptic plasticity and learning \cite{muller2016rotating}, and computation \cite{zhaoping2026conduction}. Beyond biological systems, delays have recently shown to play an important role in artificial neural networks, where communication latencies affect the coordination and efficiency of distributed computation \cite{benigno2023waves,budzinski2024exact,tavakoli2024boosting,tavakoli2025signal}.

From a theoretical and numerical perspective, delays are known to alter the dynamical behaviour of networked systems \cite{atay2003distributed,zou2013reviving,koseska2013oscillation,zou2015restoration}. They can shape pattern formation in spiking networks \cite{yanchuk2017spatio, hansen2022effect}, induce oscillation death \cite{reddy1998time,zou2009partial}, promote coherence resonance \cite{masoliver2017coherence}, trigger extreme events \cite{saha2017extreme}, and generate spontaneous traveling waves \cite{davis2021spontaneous}. More generally, time delays can change stability properties \cite{crook1997role,mihara2019stability,lee2024stability,sinha2025geometric} replacing phase-synchronized states with waves or other spatially structured activity patterns \cite{jeong2002time,budzinski2023analytical,sinha2025geometric}.

In this work, we contribute to the literature by describing how delayed excitatory coupling organizes activity in excitable networks. We show that sufficiently large transmission delays amplify irregular spiking activity, which in turn drives qualitative changes in collective behavior of the network. As coupling strength increases, the network first self-organizes into an anti-phase synchronized state and then undergoes an abrupt transition to phase synchronization. This mechanism requires only three minimal ingredients: irregular spiking activity (arising from heterogeneity, noise, or chaos), excitatory interactions, and sufficiently large delays. Importantly, the phenomenon is general and robust across network architectures, neuronal models, and sources of irregularity. Our results identify delayed excitation as a general coordinator of collective dynamics in excitable networks, and provide a mechanistic framework for understanding how temporal latencies may shape neural computation.

\begin{figure*}[thb]
    \centering
    \includegraphics[width=.95\linewidth]{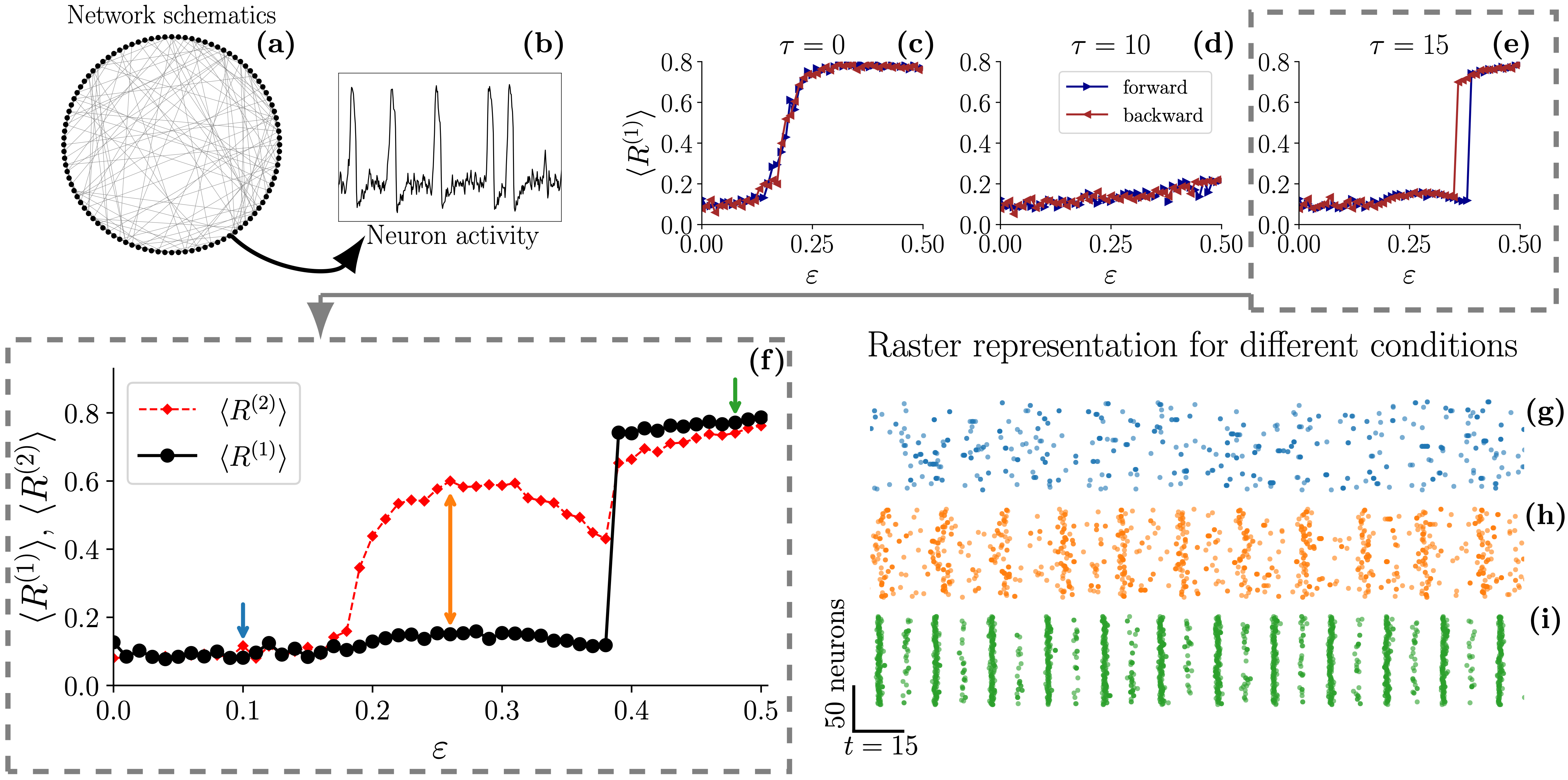}
    \caption{\textbf{Delayed connections lead to an abrupt transition to phase synchronization}. \textbf{(a)} We define our system on a random network. In this case, we consider a Watts-Strogatz network with $N = 100$, $k = 10$, and $p = 0.1$. \textbf{(b)} Neurons display irregular spiking activity. We numerically integrate the network and compute $R^{(1)}$ as a function of the coupling $\varepsilon$ as we increase and decrease $\varepsilon$ in a continued manner. \textbf{(c)}  Without delay, we observe a smooth transition to phase synchronization. \textbf{(d)} For intermediate values of $\tau$ ($\tau = 10$ here), $R^{(1)}$ remains low for the entire range of $\varepsilon$. \textbf{(e)} When the delay becomes sufficiently large ($\tau = 15$ here), we observe an abrupt transition to phase synchronization as we increase the coupling. As we decrease the coupling, we observe an abrupt transition from high to low values of $R^{(1)}$. The transition does not occur at the same $\varepsilon$ value, however, and the system displays hysteresis. \textbf{(f)} We compute $R^{(1)}$ and $R^{(2)}$ as a function of $\varepsilon$ for $\tau = 15$. We observe an increase in $R^{(2)}$ before the abrupt transition, indicating the existence of anti-phase clusters. We then compute the raster plots for illustrative values of $\varepsilon$ in this process. \textbf{(g)} For $\varepsilon = 0.1$, the network is asynchronous. \textbf{(h)} For $\varepsilon = 0.26$, when $R^{(2)}$ is high and $R^{(1)}$ is low, the network starts to displays anti-phase clusters. \textbf{(i)} After the abrupt transition ($\varepsilon = 0.47$). We note the order parameters are given by a temporal average over $10,000$ timesteps for each $\varepsilon$ value.}
    \label{fig:main}
\end{figure*}

To discuss this mechanism, we consider a network of FitzHugh-Nagumo neurons \cite{fitzhugh1961impulses} driven by white noise, following
\begin{align}\label{eq:fn_neurons}
    \dot x_i(t) &= x_i(t) - \frac{x_i^3(t)}{3} - y_i(t) + I_{i,\mathrm{coup}}(t) + D\xi_i(t) ,\\ 
    \dot y_i(t) &= c\big(x_i(t) + a - b y_i(t)\big).
\end{align}
Here, $x_{i}(t)$ is the fast variable of neuron $i$ at time $t$, which can be interpreted as the membrane potential of the neuron, $y_i(t)$ is the slow, or recovery variable, and $D\xi_i(t)$ represents an additive stochastic input, where $\xi_i(t)$ is drawn from Gaussian white noise with zero mean and unit variance, and $D$ denotes the noise intensity. We use standard parameters for the model: $a = 0.7$, $b = 0.8$, and $c = 0.1$. Due to the noise in the system, the individual behaviour is given by irregular spiking activity (Fig.~\ref{fig:main}b).

The term $I_{i,\mathrm{coup}}(t)$ represents the synaptic coupling given by \cite{destexhe1994efficient}
\begin{equation}\label{eq:coup}
    I_{i,\mathrm{coup}} = \frac{\varepsilon}{\eta}\big(x_s - x_i(t)\big)\sum_{j=1}^{N} A_{ij}\, s_j(t-\tau),
\end{equation}
where $\varepsilon$ controls the coupling strength, $\eta$ is the average degree of connectivity of the network, $x_s = 2$ is the synaptic reversal potential, $A_{ij}$ represents the elements of the adjacency matrix, and $\tau$ represents the time delay in the interactions. The function $s_i(t)$ models the synaptic interaction and is defined as
\begin{equation}\label{eq:synaptic}
    s_i(t) = \Big(1 + e^{-\lambda x_i(t)}\Big)^{-1},
\end{equation}
where $\lambda = 30$ determines the steepness of the sigmoidal response. With this, the information from the presynaptic neuron is transmitted to the postsynaptic neuron after a finite propagation time $\tau$, which we call the delay (see Supplementary Fig.~S1 for an illustration).

To explore how delayed connections can coordinate collective dynamics, we 
consider a complex network. In this case, the adjacency matrix $\bm{A}$ is constructed using the Watts-Strogatz model~\cite{watts1998collective} with $N=100$ neurons, each initially connected to its $k=10$ nearest neighbors and then rewired with probability $p = 0.1$ which generates an average of degree of connectivity $\eta \approx 20$ (a schematic representation of the network is displayed in Fig.~\ref{fig:main}a). To characterize the collective dynamics of the system, we compute the $m$-th moment of the order parameter \cite{sepulchre2007stabilization}
\begin{equation}
    R^{(m)}(t) = \left | \frac{1}{N} \sum_{j=1}^N e^{\i m \varphi_j(t)}\right|,
    \label{eq:order_parameter}
\end{equation}
where $R^{(m)} \in [0, 1]$ is the order parameter of order $m \in \N^*$, $\i = \sqrt{-1}$ is the complex unit, and $\varphi_j(t)$ is the instantaneous phase of neuron $j$ at time $t$. We note the phase is defined using an event-based approach \cite{ivanchenko2004phase}, in which events are identified by threshold crossings of the dynamical variable $x_j$ at $x^*=0$ (see Supplementary Fig.~S2 for details). For $m = 1$, we obtain the classical Kuramoto order parameter \cite{kuramoto1975self} that measures phase synchronization ($R^{(1)} = 1$). For $m = 2$, the order parameter is suited to detecting anti-phase synchronization ($R^{(2)} = 1$) -- see Supplementary Fig.~S3 for examples.

Figures \ref{fig:main}c-e show the evolution of the time-averaged phase synchronization order parameter $\langle R^{(1)}\rangle$ as a function of the coupling strength $\varepsilon$ for different values of the delay $\tau$. Here, we employ a continuation protocol in which the coupling is updated according to $\varepsilon \rightarrow \varepsilon + \delta\varepsilon$. The coupling strength is swept in both the forward ($\delta\varepsilon=0.01$) and backward ($\delta\varepsilon=-0.01$) directions (see Supplemental Material, Sec.~II C for details). Throughout this procedure, the system is not reinitialized; instead, the final state obtained at each value of $\varepsilon$ is used as the initial condition for the subsequent step, ensuring that the dynamics remain on the same attractor whenever possible. When the delay is null $\tau = 0$, the system undergoes a smooth transition from an asynchronous to a phase synchronized state (Fig.~\ref{fig:main}c). For $\tau=10$, the presence of delay suppresses phase synchronization, keeping the network in a state with low $\langle R^{(1)}\rangle$ (Fig.~\ref{fig:main}d). When the delay is larger ($\tau = 15$, for example), the system exhibits an abrupt transition from incoherence to synchronization, followed by a pronounced hysteresis loop arising from the different forward and backward continuation branches (Fig.~\ref{fig:main}e).

We then focus on this case ($\tau = 15$, considering the forward direction of increasing coupling) and analyze the first and second moments of the order parameter (black and red lines in Fig.~\ref{fig:main}f). As the coupling strength increases, the system initially exhibits cluster synchronization, in which a fraction of the neurons spike together while the remaining neurons spike in between the dominant group. This configuration results in an partial anti-phase synchronized state, characterized by a high value of $R^{(2)}$ and a low value of $R^{(1)}$. Upon further increase of the coupling, the system undergoes an abrupt transition to partial phase synchronization, where both $R^{(2)}$ and $R^{(1)}$ attain high values, and full phase synchronization is not reached due to noise. 

The spiking activity of the system is shown in three illustrative examples, for different coupling strength values $\varepsilon$ (blue, orange, and green arrows, Fig.~\ref{fig:main}f). Figures \ref{fig:main}g-i display raster plots, where each dot represents a spike. When the coupling is low, the spiking activity is asynchronous (Fig.~\ref{fig:main}g). With intermediate coupling, two clusters emerge, representing an initial stage of a partial anti-phase synchronization (Fig.~\ref{fig:main}h). After the abrupt transition, we observe that most neurons spike synchronously, while a few remain in anti-phase (Fig.~\ref{fig:main}i). This small anti-phase cluster occurs due to the delayed interacting mechanism we explain below (see Supplementary Fig.~S4). We note that the same qualitative behaviour occurs in a variety of circumstances, including different network architectures and in neuronal models with distinct dynamics (excitable with noise and chaotic) -- Supplementary Figs.~S5-S7 show further examples.

\begin{figure}[t]
    \centering
    \includegraphics[width=\linewidth]{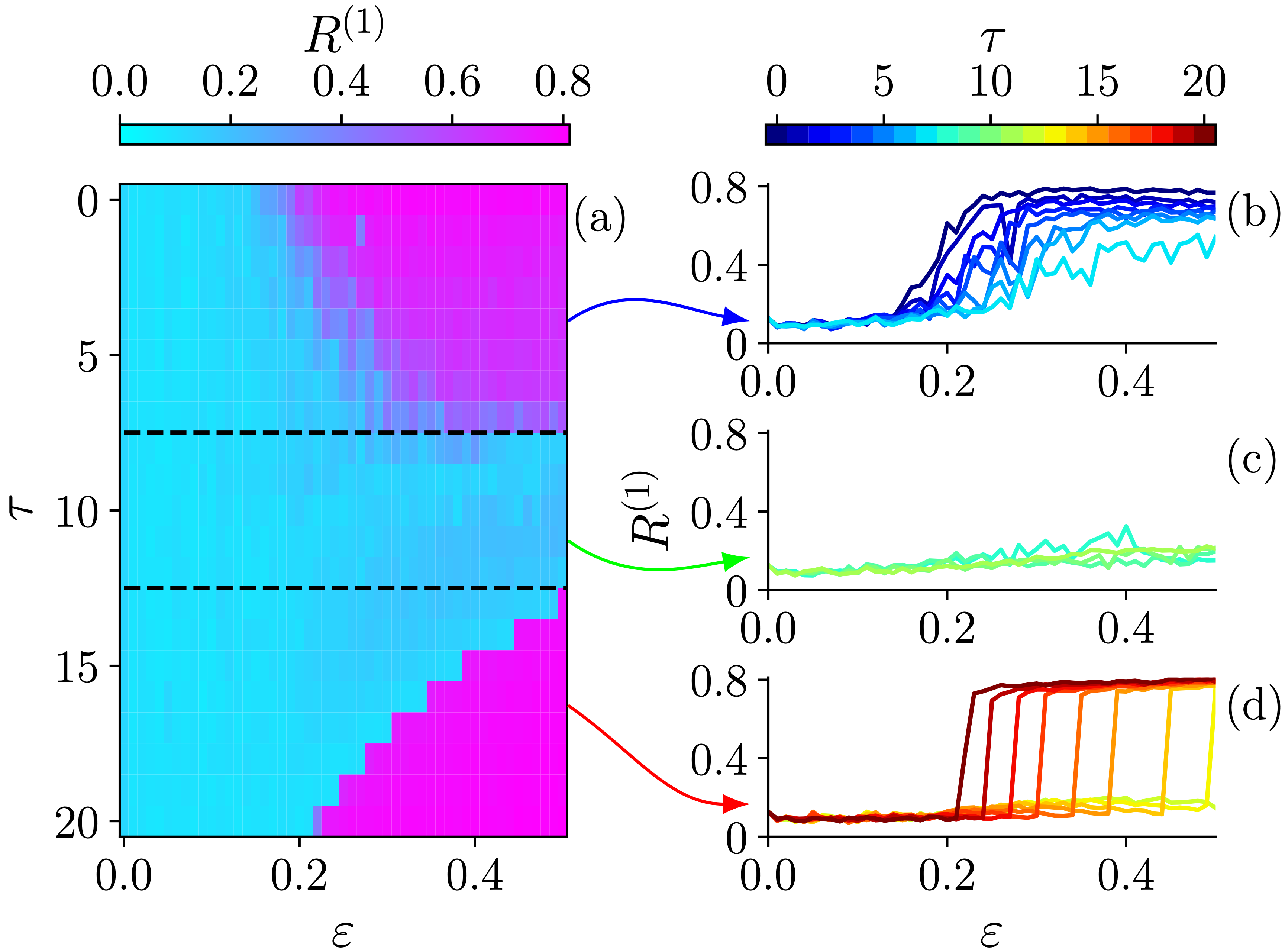}
    \caption{\textbf{Abrupt transition occurs for a wide range of delays.} We consider the same network as studied in Fig.~\ref{fig:main} and compute the temporal average of first moment of the order parameter ($\langle R^{(1)}\rangle$) as a function of $\varepsilon$ for different $\tau$ values (a). We adopt a continued increase of the coupling strength. We can observe a non-monotonic behaviour of $\langle R^{(1)}\rangle$ as a function of $\tau$. For low values of $\tau$  (b), the network displays a smooth transition to phase synchronization. For intermediate values of $\tau$, $\langle R^{(1)}\rangle$ remains low for the range of $\varepsilon$ considered here  (c). If the delay is large enough  (d), however, we observe an abrupt transition to phase synchronization. We also note that, for these cases, the larger the delay, the sooner the network transitions.}
    \label{fig:multiple}
\end{figure}

The results therefore, show that delays can have markedly different effects on activity patterns: depending on the delay value, the system may exhibit anti-phase or clustered synchronization or undergo abrupt transitions to strongly phase-synchronized states. These effects persist over a wide range of delay values. Figure \ref{fig:multiple}(a) shows the phase synchronization measure $\langle R^{(1)}\rangle$ in color-codes from cyan ($\langle R^{(1)} \rangle\approx 0$) to magenta ($\langle R^{(1)} \rangle\approx 0.8$) as a function of $\varepsilon$ (x-axis) and the delay time $\tau$ (y-axis). For small delays, $0 \leq \tau \leq 7$, the system displays a smooth transition to phase synchronization; however, the transition becomes progressively harder to access, requiring larger coupling strengths (Fig.~\ref{fig:multiple}(b) for details). For intermediate delays, $7 < \tau \leq 13$, phase synchronization is suppressed, entering in clustering or anti-phase synchronization regime (Fig.~\ref{fig:multiple}(c)). For larger delays, $\tau > 13$, the system undergoes the abrupt transition to phase synchronization. An interesting phenomenon is that longer delays reduce the critical coupling strength $\varepsilon^*$ at which the networks transition to the phase synchronized state (Fig.~\ref{fig:multiple}(d)).

\begin{figure}[t]
    \centering
    \includegraphics[width=\linewidth]{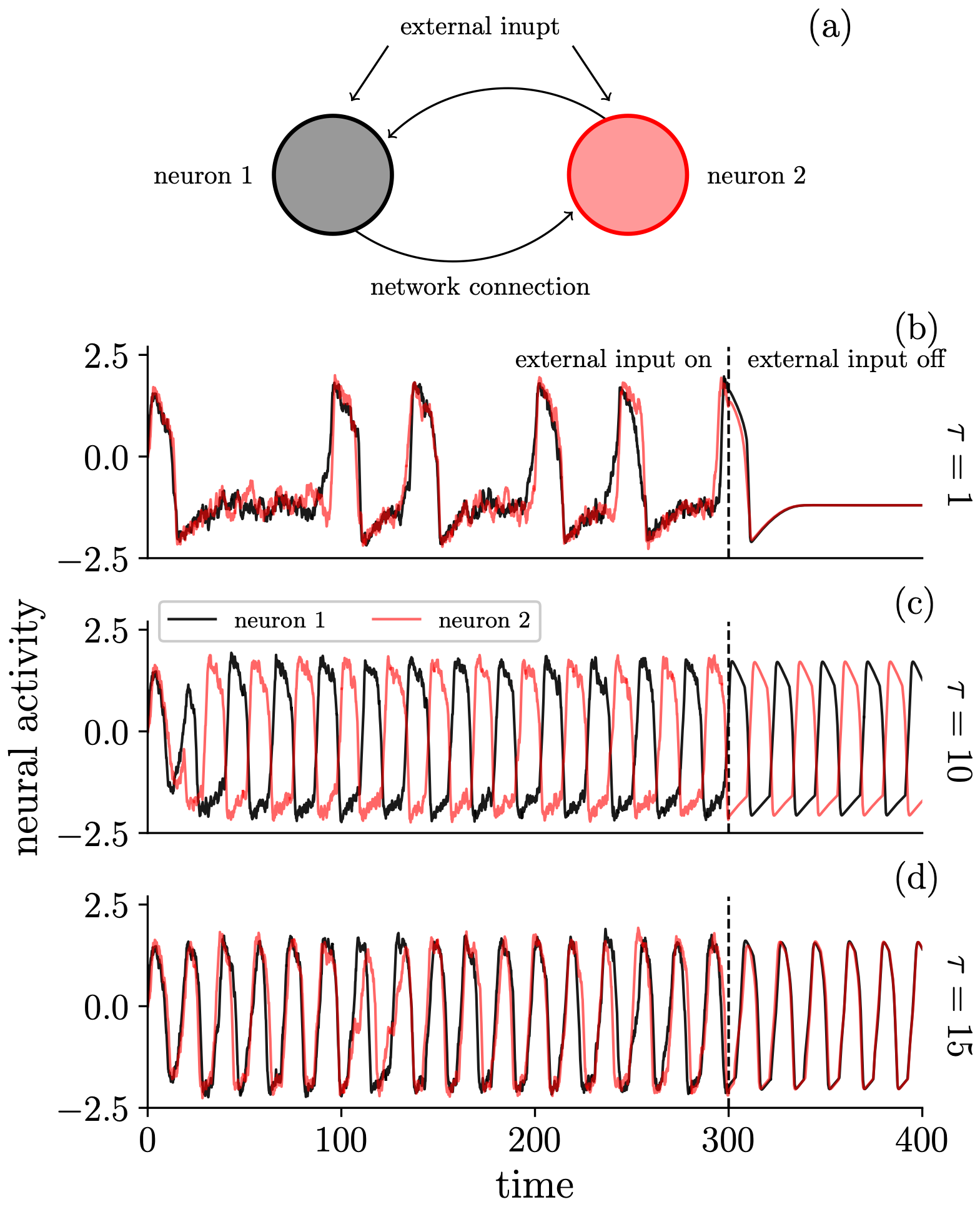}
    \caption{\textbf{Delayed coupling coordinates and sustains spiking activity.} \textbf{(a)} We consider a simplified network composed of two interacting neurons. Each neuron receives external inputs due to the stochastic noise ($D = 0.2$) and is also coupled to the other neuron. We consider a fixed value of coupling $\varepsilon=0.4$, and vary the value of the delay $\tau$. The system is initialized in a synchronized state and integrated for $300$ timesteps, where after that, we turn the external input off. \textbf{(b)} For $\tau=1$, the network is not able to maintain the synchronized firing, and after the external input is turned off, both neurons relax to a stable equilibrium with no activity. \textbf{(c)} For $\tau=10$, the network displays an anti-phase spiking activity, which is sustained even after the external input is off. \textbf{(d)} For $\tau = 15$, the network remains with synchronous firing, even after the external input is off.}
    \label{fig:mechanism}
\end{figure}

These antithetical effects of delay arise from a simple overarching mechanism. To understand it, we first note that the delay interacts with the coupling to create stable self-sustained oscillations. For simplicity, we illustrate this for two coupled neurons (Fig.~\ref{fig:mechanism}a) with fixed coupling $\varepsilon =0.4$. Similarly to Fig~\ref{fig:multiple}, this coupled system has three regimes: occasional irregular spiking for low delay, regular anti-phase spiking for intermediate delay, and regular in-phase spiking for longer delay. This is seen in Figs.~\ref{fig:mechanism}b-d, from time $t = 0$ to $t = 300$. At $t = 300$ the stochastic drive is removed, leaving the system purely deterministic. For a small delay  ($\tau = 1$, Fig.~\ref{fig:mechanism}b), the trajectories cease to oscillate and converge onto a fixed point with no activity. Crucially, for larger delays the system describes self-sustained oscillations, either in anti-phase for intermediate ($\tau = 10$, Fig.~\ref{fig:mechanism}c) or in-phase for longer delay ($\tau = 15$, Fig.~\ref{fig:mechanism}d). Therefore, the same strength of coupling may or may not create self-sustained oscillations depending on the length of the delay. The same phenomenon occurs in the inverse situation, with fixed delay and varying coupling: the self-sustained oscillations emerge for critical values of the coupling strength (Supplementary Fig.~S8). When they emerge, the oscillations, corresponding to stable limit cycles, dominate the state space, explaining the abrupt transitions seen at the network levels. Further, we note this phenomenon is robust to heterogeneity in the delays (Supplementary Fig.~S9).

To understand how these attractors emerge, we consider a coupling strength that is large enough so that neuron 1's pulse can elicit a spike in neuron 2's, originally at rest on a fixed point. For a small delay, 2 will spike soon after 1, and will influence 1 soon after that. At this point, 1 will still be performing its excitation. The influence from neuron 2 might delay the excitation of 1 and bring the two spikes closer in phase. But it will otherwise be lost, with no new spike generated. If, however, the delay is sufficiently long, 1's pulse will only reach 2 once 1 has finished its excitation and is close to rest again. At this point, 2 will spike and send its pulse to 1. This pulse will only reach 1 once 2 is again at rest and will elicit firing in 1. This behavior repeats, with one neuron spiking while the other rests. This creates an anti-phase pattern. Now consider that by chance, 1 and 2 fire at similar times (due to an initial condition or to noise). If the delay is large enough, the pulse from one neuron arrives at the other when they are again close to rest, and will cause both to fire again synchronously. In this case, the delay must be large enough that the excitation reaches a state where the pulse can elicit further spikes. In the anti-phase case, the delay can be smaller, since the trajectories will already be displaying an excitation when the other neuron`s pulse is sent. We note this is consistent with previous work exploring the role of delays in excitable systems \cite{buric2003dynamics,dahlem2009dynamics}, which we can now leverage to explain the collective behavior and organization at the network level. This mechanism has a large explanatory power, describing how (i) the anti-phase oscillation emerges for smaller $\tau$ while the in-phase oscillation does so for longer $\tau$, (ii) coexistence of the two attractors, which are reached depending on initial conditions or noise, (iii) the existence of critical values for delay given fixed coupling, (iv) existence of critical values for coupling given fixed delay, (v) the generality of behavior, which does not require specific ingredients of the neuronal model or coupling.

In this work, we have highlighted a network mechanism through which delayed excitatory connections coordinate spiking activity and induce abrupt transitions to synchronization in excitable networks. Our results provide direct evidence that delayed excitatory interactions can structure spiking activity by first promoting anti-phase firing patterns and, with increasing coupling, triggering an abrupt transition to in-phase synchronization. We show that in networks with delayed excitatory interactions and irregular spiking activity, the delay can amplify and coordinate noise-induced neural fluctuations, providing a mechanism for the emergence of collective dynamics and abrupt synchronization transitions. Importantly, this mechanism is largely independent of initial conditions, specific network connectivity, neuronal model, or presence of noise (Supplementary Figs.~S5–S7), being therefore quite general. It also advances a class of recently described trapping mechanisms by which coupling between units interacts with transients in their individual dynamics to generate attractors and possibly multistability in the networks \cite{medeiros2018boundaries,medeiros2019state,medeiros2021the,contreras2023scale,rossi2025transients}.

Our findings are consistent with previous studies on explosive synchronization, originally reported in oscillator networks \cite{gomez2011explosive} and later observed in spiking neuron networks \cite{boaretto2019mechanism}. Here we identify delayed excitable interactions as a key ingredient capable of coordinating collective dynamics in spiking systems to generate explosive synchronization. Such mechanisms may be relevant for understanding neural function. For example, long-range connections have been proposed to coordinate neural activity \cite{battaglia2007temporal} in processes such as working memory \cite{mejias2022mechanisms}; interactions between timescales and external stimulation have been shown to profoundly change synchronization in somatosensory neural paths \cite{sagalajev2024absence}; and delays have recently been shown to enhance the computational capacity of neural networks \cite{budzinski2024exact,tavakoli2024boosting,tavakoli2025signal,peddinti2015time,sun2023learnable}. More broadly, the mechanism described here provides a simple and robust route by which delays can control collective dynamics and generate abrupt synchronization transitions in excitable systems. Future work could explore this mechanism in networks with distance-dependent delays, where it may give rise to spatially structured patterns of synchronization and desynchronization.

\section*{Acknowledgments}
B.R.R.B. and E.E.N.M. are supported by the Brazilian São Paulo Research Foundation (FAPESP), Proc. 2021/09839-0, 2023/16273-8, and 2024/05700-5 | ``Dinâmica Não Linear'' and Conselho Nacional de Desenvolvimento Cientifico e Tecnológico (CNPq) through fellowship and Grant No. 150376/2026-0 | ``INCT NeuroComp''.  R.C.B acknowledges the support of the Natural Sciences and Engineering Research Council of Canada (NSERC) [grants RGPIN-2026-05758 and DGECR-2026-00066] and the support of the Canada Research Chairs program. K.L.R. acknowledges support by the Max Planck Society.

\section*{Code availability}
An open-source repository with codes used in this work will be available at \href{https://github.com/budzinskilab}{\textcolor{Cerulean}{github.com/budzinskilab}}.


%

\end{document}